\def\wb{{\widetilde{\beta} }}
\def\rb{{r_{\!_{\wb}}}}
\def\DI{{D_{\text{I}}}}
\def\Dh{{D_{\text{H}}}}

\documentclass[aps,prappl,twocolumn,groupedaddress,showkeys,showpacs,superscriptaddress,floatfix,longbibliography]{revtex4-1}
\usepackage{epsfig}
\usepackage{multirow}
\usepackage{amsmath, amssymb,mathtools}
\usepackage{color}
\usepackage{graphicx}
\usepackage{dcolumn}
\usepackage{bm}
\usepackage{subfigure}
\usepackage[utf8]{inputenc}
\usepackage[T1]{fontenc}
\usepackage{mathptmx}
\usepackage{tikz}

\definecolor{SB-color}{named}{blue}

\definecolor{CG-color}{named}{red}

\begin{document}

\title{A cryogenic memory element based on an anomalous Josephson junction}

\author{C. Guarcello}
\affiliation{Centro de Física de Materiales, Centro Mixto CSIC-UPV/EHU, Paseo Manuel de Lardizabal 5, 20018 San Sebastián, Spain}
\affiliation{Dipartimento di Fisica ``E.R. Caianiello'', Universit\`a di Salerno, Via Giovanni Paolo II, 132, I-84084 Fisciano (SA), Italy}
\author{F.S. Bergeret}
\affiliation{Centro de Física de Materiales, Centro Mixto CSIC-UPV/EHU, Paseo Manuel de Lardizabal 5, 20018 San Sebastián, Spain}
\affiliation{Donostia International Physics Center, Paseo Manuel de Lardizabal 4, 20018 San Sebastián, Spain}


\begin{abstract}
We propose a non-volatile memory element based on a lateral ferromagnetic Josephson junction with spin-orbit coupling and out-of-plane magnetization. The interplay between the latter and the intrinsic exchange field of the ferromagnet leads to a magnetoelectric effect that couples the charge current through the junction and its magnetization, such that by applying a 
current pulse the direction of the magnetic moment in F can be switched. The two memory states are encoded in the direction of the out-of-plane magnetization. With the aim to determine the optimal working temperature for the memory element, we explore the noise-induced effects on the averaged stationary magnetization by taking into account thermal fluctuations affecting both the Josephson phase and the magnetic moment dynamics. We investigate the switching process as a function of intrinsic parameters of the ferromagnet, such as the Gilbert damping and strength of the spin-orbit coupling, and proposed a non-destructive readout scheme based on a dc-SQUID. Additionally, we analyze a way to protect the memory state from external perturbations by voltage gating in systems with a both linear-in-momentum Rashba and Dresselhaus spin-orbit coupling. 

\end{abstract}

\maketitle

\section{Introduction}
\label{Introduction}\vskip-0.2cm

Superconducting electronics is suggested as playing an important role in the search of ultra-low-power computers~\cite{herr2011ultra,nishijima2013superconductivity,Lik91,Muk11,Hol13}.
One of the key challenges towards this objective is the fabrication of a reliable and scalable cryogenic memory architecture. Superconductor-ferromagnet-superconductor (SFS) junctions are promising structures suggested for such memories~\cite{Feo10,Rya12,Gol12,Bae14,Gol15,Gin16,Bet17,Day18,DeS18,Pra19}. 
Indeed, the interplay between the intrinsic exchange field and the induced superconductivity in the ferromagnet 
leads to the so-called $\pi$-junction, {\it i.e.}, a Josephson junction exhibiting an intrinsic $\pi$-phase shift in its ground state. 
Vertical ferromagnetic multilayer structures are being used as Josephson magnetic memories. The two logic states of these memories usually correspond to states with different relative orientation of magnetic layers, that in turn determines whether the junction is in the $0$- or $\pi$-state. Readout schemes are commonly based on distinguish between resistive and non-resistive states. 

\begin{figure}[b!!]
\centering
\includegraphics[width=\columnwidth]{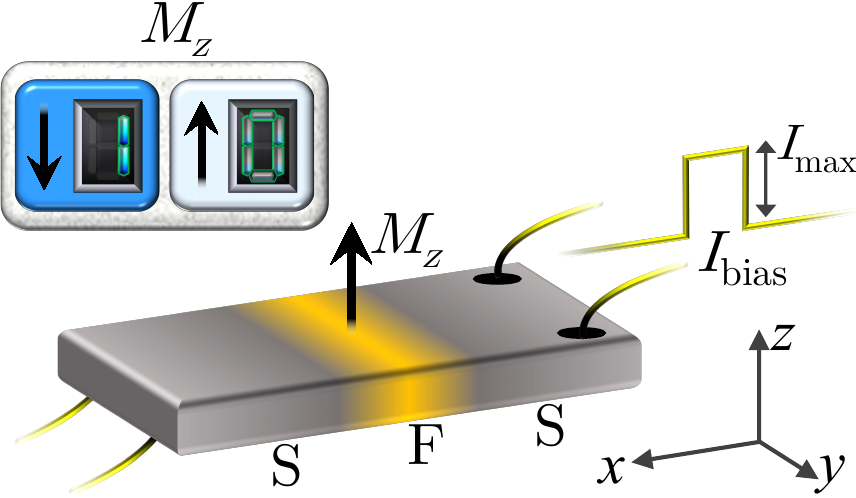}
\caption{S/F/S Josephson junction driven by a rectangular bias current pulses, $I_{\text{bias}}$, with amplitude $I_{\text{max}}$. The $z$-component of the magnetization, $M_z$, is the observable used to define the logic memory states 0 and 1.} 
\label{Figure01}
\end{figure}

Here, we delve into the alternative idea, initially suggested in Ref.~\cite{Shu17}, of designing a cryogenic memory element based on a ferromagnetic anomalous Josephson junction, usually called $\varphi_0$-junction~\cite{Buz08}. It consists of a SFS Josephson junction with a Rashba-like spin-orbit coupling (SOC). Its ground state corresponds to a finite phase shift in its current-phase-relation $0<\varphi_0<\pi$. Such anomalous phase has been recently detected experimentally in hybrid Josephson junctions fabricated with the topological insulator Bi$_2$Se$_3$~\cite{Ass19} and Al/InAs heterostructures~\cite{May19} and nanowires~\cite{Str20}. Both materials have strong spin-orbit coupling and in those experiments, time reversal is broken by an external magnetic field that acts as a Zeeman field. 
Here, the memory element is a Josephson junction with a ferromagnetic link, thus in principle time-reversal is broken intrinsically by the exchange field. As demonstrated theoretically, in these junctions the magnetization of F can be controlled by passing an electric current through the device~\cite{Wai02,Kon09,Lin11,Shu17,Bob18,Shu18,Maz19}. We discuss the use of such a junction as a memory element with the information encoded in the magnetization direction of the F layer. An important issue for the memory element is the effects stemming from the unavoidable thermal fluctuations on both phase and magnetization dynamics. Therefore we present an exhaustive analysis of the noisy dynamics of a current-biased SFS Josephson junction, see Fig.~\ref{Figure01}, considering the influence of stochastic thermal fluctuations. Notably, in order to preserve gauge-invariance the numerical model that we use to describe the phase dynamics includes also the time derivative of the anomalous phase, according to Ref.~\onlinecite{Rab19}. 
We explore the current-induced magnetic bistability, in order to define two well-distinguishable logic states, and investigate the robustness of such memory against noise-induced effects. We also suggest a suitable non-invasive readout scheme based on a current-controlled superconducting quantum interference device (SQUID), which does not require an additional magnetic flux to set the optimal operating point. Moreover, we discuss the intriguing possibility of effectively shielding the memory state by voltage gating, in a device formed by a ferromagnetic layer with a linear-in-momentum Dresselhaus spin-orbit coupling term.

The work is organized as follows: In Sec.~\ref{LLG+RSJ}, we present the theoretical model used to describe the time evolution of both the magnetic moment and the Josephson phase of a current-driven SFS junction. In Sec.~\ref{deterministicanalysis} we discuss the magnetic configuration of the junction after applying a current pulse. Specifically we determine its stationary magnetization, as a function of the Gilbert damping parameter and the SOC strength. In Sec.~\ref{noiseeffects}, we focus on the effect of stochastic thermal fluctuations on both the phase and the magnetization dynamics. In Sec.~\ref{GenericSOC}, we explore how a generic Rashba-Dresselhaus type SOC can affect the stationary magnetization. In Sec.~\ref{Readout}, we discuss a feasible readout scheme based on a dc-SQUID magnetometer. In Sec.~\ref{Conclusions} we present our conclusions. 

\section{The model}
\label{LLG+RSJ}\vskip-0.2cm

We consider a similar setup as the one studied in Ref.~\cite{Kon09}. It consists of a SFS junction in which the ferromagnet is a thin film with an out-of-plane magnetic anisotropy and a Rashba-like SOC, see Fig.~\ref{Figure01}. 

Because of the interplay between the exchange field and the SOC, the current-phase relation of the SFS Josephson junction reads $I_{\varphi}=I_c\sin(\varphi-\varphi_0)$, where $I_c$ is the critical current of the junction, $\varphi$ is the Josephson phase difference, and $\varphi_0$ is the so-called anomalous phase shift. The latter depends on several parameters of the system, such us the Rashba coefficient $\alpha$~\cite{Ras60,Byc84}, the transparency of S/F interfaces, the spin relaxation, and the disorder degree of the junction. The exact dependence on these parameters is not as crucial as the geometry of the device. If we consider a two-dimensional SOC with momenta in the plane of the F film, and the charge current flows in $x$-direction, then the phase shift $\varphi_0$ is proportional to the $y$-component of the magnetic moment according to~\cite{Buz08,Kon09,Kon15,Ber15}
\begin{equation}\label{phi0}
\varphi_0=r\frac{M_y}{M},
\end{equation}
where $M=\sqrt{M_x^2+M_y^2+M_z^2}$ is the modulus of the magnetization vector, and the parameter $r$ quantifies the SOC strength and contains the $\alpha$-dependence. The specific dependence of $r$ on the SOC will be discussed in Sec.~\ref{noiseeffects}, when we consider the coexistence of both Rashba and Dresselhaus SOC contributions and their impact on the magnetization dynamics.

As discussed in Refs.~\cite{Kon09,Shu17,Shu18,Ata19}, Eq.~\eqref{phi0} establishes a direct coupling between the magnetic moment and the supercurrent. The time evolution of the magnetization can be described in terms of the Landau–Lifshitz–Gilbert (LLG) equation~\cite{Lan35,Gil04}
\begin{equation}\label{LLG}
\frac{d\textbf{M}}{d\tau}=-g_r\textbf{M}\times\textbf{H}_{\text{eff}}+\frac{\gamma}{M}\left ( \textbf{M}\times\frac{d\textbf{M}}{d\tau} \right ),
\end{equation}
where $g_r$ denotes the gyromagnetic ratio. The first term on the right-hand side describes the precession motion around $\textbf{H}_{\text{eff}}$, whereas the second term describes the dissipation, which is accounted by the phenomenological dimensionless Gilbert damping parameter $\gamma$. The $i$-th component of the effective field $\textbf{H}_{\text{eff}}$, with $i=x,y,z$, can be calculated as~\cite{Lif90}
\begin{equation}\label{EffectiveField_FreeEnergy}
H_{\text{eff},i}=-\frac{1}{\Omega}\frac{\partial \mathcal{F}}{\partial M_i},
\end{equation}
where $\Omega$ is the volume of the F layer, and $\mathcal{F}$ is the free energy of the junction, which can be written as
\begin{equation}\label{FreeEnergy}
\mathcal{F}=-E_J\varphi I_{bias}+E_s(\varphi,\varphi_0)+E_M.
\end{equation}
Here, $E_J=\Phi_0I_c/(2\pi)$ with $\Phi_0$ being the flux quantum, $I_{bias}$ is the external current in units of $I_c$, $E_s(\varphi,\varphi_0)=E_J[1-\cos(\varphi-\varphi_0)]$, and $E_M=-\frac{\mathcal{K}\Omega}{2}\left ( \frac{M_z}{M} \right )^2$ is the magnetic energy depending on the anisotropy constant $\mathcal{K}$. 
The ratio between the energy scales of the system will be indicated by the parameter $\varepsilon =E_J/(\mathcal{K}\Omega)$.

The effective magnetic field calculated from Eq.~\eqref{EffectiveField_FreeEnergy} reads
\begin{equation}\label{effectivefield}
\textbf{H}_{\text{eff}}=\frac{\mathcal{K}}{M}\left [ \varepsilon r \sin\left ( \varphi-rm_y \right )\hat{y}+m_z\hat{z} \right ],
\end{equation}
with $m_{x,y,z}=M_{x,y,z}/M$.

Since the normalized components of the magnetization satisfy the condition $m_x^2+m_y^2+m_z^2=1$, 
it is convenient to write the LLG equations in spherical coordinates~\cite{Rom14}, so that the normalized components of the magnetization can be expressed in terms of the polar and azimuthal angles $\theta$ and $\phi$ as
\begin{eqnarray}\label{Msphericalcoord}\nonumber
m_x(\tau)&=&\sin\theta(\tau)\cos\phi(\tau)\\
m_y(\tau)&=&\sin\theta(\tau)\sin\phi(\tau)\\\nonumber
m_z(\tau)&=&\cos\theta(\tau).
\end{eqnarray}
By normalizing the time to the inverse of the ferromagnetic resonance frequency $\omega_F=g_r \mathcal{K}/M$, that is $t=\omega_F\tau$, the LLG equations in spherical coordinates reduce to the following two coupled equations~\cite{Rom14}
\begin{eqnarray}\label{LLGsphericalcoord_a}
\frac{\mathrm{d} \theta}{\mathrm{d}t}=&&\frac{1}{1+\gamma^2}\left ( \widetilde{H}_{\text{eff},\phi}+\gamma \widetilde{H}_{\text{eff},\theta}\right )\\\label{LLGsphericalcoord_b}
\sin \theta \frac{\mathrm{d} \phi}{\mathrm{d} t}=&&\frac{1}{1+\gamma^2}\left (\gamma \widetilde{H}_{\text{eff},\phi}- \widetilde{H}_{\text{eff},\theta} \right ),
\end{eqnarray}
where the $\theta$ and $\phi$ components of the normalized effective field are defined as
\begin{eqnarray}\label{AngleFieldsEffective-a}
&&\widetilde{H}_{\text{eff},\theta}=\varepsilon r\sin(\varphi-rm_y)\cos\theta\sin\phi-m_z\sin\theta\qquad\\\label{AngleFieldsEffective-b}
&&\widetilde{H}_{\text{eff},\phi}=\varepsilon r\sin(\varphi-rm_y)\cos\phi.
\end{eqnarray}

Next, we discuss how the Josephson phase $\varphi(t)$ of a magnetic junction responds to a driving bias current. The dynamics of an overdamped SFS Josephson junction can be described through the modified resistively shunted junction (RSJ) model~\cite{Bar82,Rab19} generalized to include the anomalous phase shift $\varphi_0=rm_y$. The phase dynamics is described by the equation:
\begin{equation}\label{RSJ}
\frac{d\varphi}{dt}=\omega\left [ I_{bias}(t)-\sin\left ( \varphi-rm_y \right )\right ] +r\frac{d m_y}{dt},
\end{equation}
where the time is still normalized to the inverse of the ferromagnetic resonance frequency and $\omega=\omega_J/\omega_F$, with $\omega_J=2\pi I_c R/\Phi_0$ being the characteristic frequency of the junction~\cite{Bar82} ($R$ is the normal-state resistance of the device). 

Last term in the right-hand side of Eq.~\eqref{RSJ} stems from the time derivative of the anomalous phase, $\dot{\varphi}_0$, {\it cf.} Eq.~\eqref{phi0}, and was ignored in Refs.~\cite{Kon09,Shu17,Shu18}, but has to be included in order to preserve gauge-invariance~\cite{Rab19}. The detailed derivation of this term can be found in the Supplemental Material of Ref.~\onlinecite{Rab19}. 
\begin{figure}[t!!]
\centering
\includegraphics[width=\columnwidth]{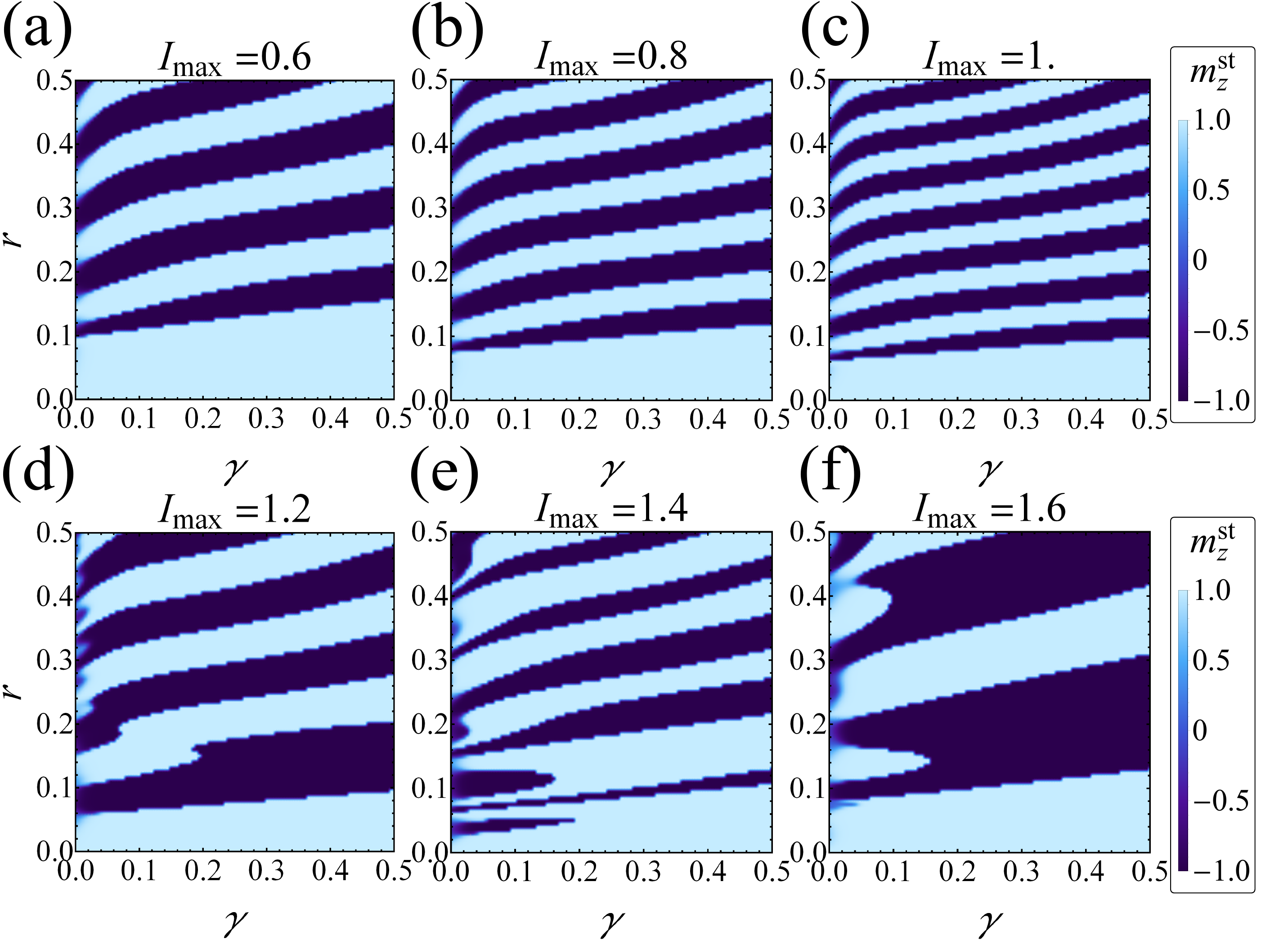}
\caption{Stationary magnetization, $m_z^{\text{st}}$, as a function of $r$ and $\gamma$, at different values of the amplitude $I_{\text{max}}$ of the current pulse, in the absence of noise fluctuations. The other parameters are: $\varepsilon =10$, $\omega=1$, $\sigma=5$, and $m_z(t=0)=+1$.} 
\label{Figure02}
\end{figure}

We assume that the SFS junction is driven by a rectangular current pulse, $I_{bias}$, centered at $t_c$: 
\begin{equation}\label{currentpulse}
I_{bias}(t)=\left\{\begin{matrix}
I_{\text{max}},&& t_c-\sigma\leq t\leq t_c+\sigma\\
0,&&\text{elsewhere}.
\end{matrix}\right.
\end{equation}
Here, $\sigma$ is the width and $I_{\text{max}}$ is the intensity, in units of $I_c$, of the pulse, so that a value $I_{\text{max}}>1$ indicates a bias current larger than the critical value.

Because of the magnetoelectric effect in a $\varphi_0$-junction, the charge current induces an in-plane magnetic moment~\cite{Ede95,Ede05,Mal08,Ber15,Kon15,Bob17}, which in turn acts as a torque on the out-of plane magnetization of the F layer and eventually leads to its switching~\cite{Kon09,Shu17}. 

In the next sections we search for an optimal combination of system parameters to induce the magnetization reversal. 
Specifically, we explore the response of the magnetization by varying $\gamma$ and $r$ in suitable ranges, whereas the energy and timescales ratios $\varepsilon$ and $\omega$, are fixed.
The energy ratio $\varepsilon$ ranges from $\varepsilon\sim100$~\cite{Kon09} in systems with weak magnetic anisotropy, to $\varepsilon\sim1$ for stronger anisotropy~\cite{Shu19}. In our calculation we choose an intermediate value $\varepsilon =10$. The typical ferromagnet resonance frequency is $\omega_F\simeq10\;\text{GHz}$, while the characteristic Josephson frequency, usually of the order of gigahertz, may be tuned experimentally. Therefore we choose $\omega=1$. As long as the injected bias current is below the critical value, the results discussed in this work are only weakly affected by the value of $\omega$. In contrast, if $I_{bias}>1$ the magnetic switching would become more unlikely as $\omega$ increases. In particular, for $\omega\gg1$ the torque exerted by the Josephson current oscillates very fast, in comparison with the timescale of the magnetization~\cite{Lin11}. This means that the magnetization would experience an effective torque averaged over many oscillations, which results in a small contribution due to a partial cancellation of the net torque.  

At $t=0$ we assume that the F magnetization points towards the $z$-direction, that is $\textbf{M}=(0,0,1)$. With this initial values we solve  Eqs.~\eqref{LLGsphericalcoord_a}-\eqref{LLGsphericalcoord_b} and~\eqref{RSJ} self-consistently for different values of the parameters. From the solution we determine the magnetization direction after the current pulse.

\section{The deterministic analysis}
\label{deterministicanalysis}\vskip-0.2cm

We first neglect the effect of thermal noise, as in Refs.~\cite{Shu17,Shu18}, and explore the magnetic switching of the junction. 

The overall behavior of the stationary magnetization $m_z^{\text{st}}$, namely, the value of $m_z$ at the time $t=t_{\text{max}}=100$, as a function of both the Gilbert damping parameter, $\gamma$, and the SOC strength, $r$, at different current pulse intensities $I_{\text{max}}$ and $\sigma=5$, is summarized in Fig.~\ref{Figure02}. We note that each contour plot is characterized by a dark fringes pattern, namely, we observe regions of the $(\gamma,r)$ parametric space in which the magnetization reversal systematically occurs, {\it i.e.}, in which $m_z^{\text{st}}=-1$. 
This means, for instance, that by increasing $r$ at a fixed $\gamma$ we observe a sequence of $m_z^{\text{st}}=+1$ and $m_z^{\text{st}}=-1$ values.

\begin{figure}[t!!]
\centering
\includegraphics[width=\columnwidth]{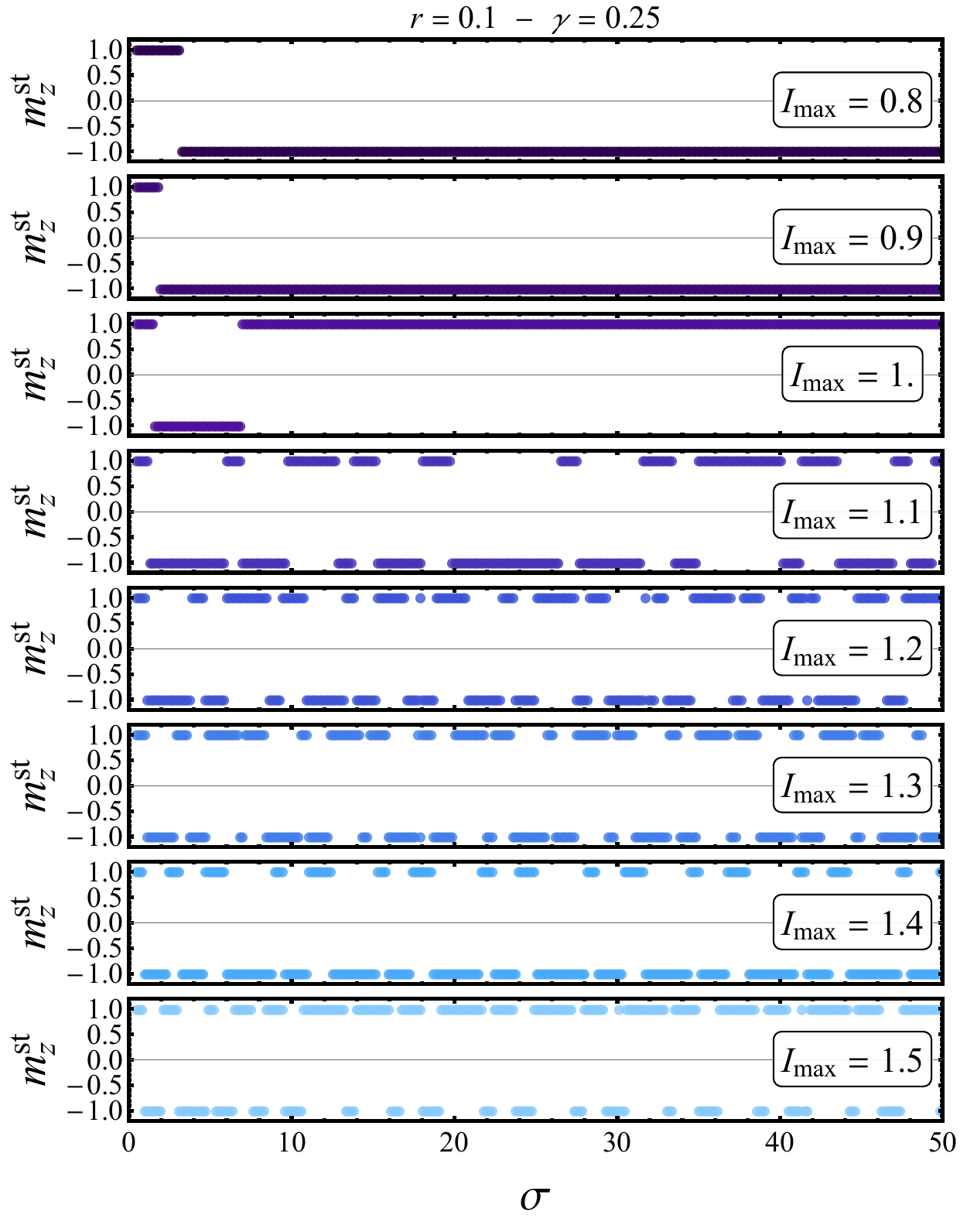}
\caption{Stationary magnetization, $m_z^{\text{st}}$, as a function of the current pulse width, $\sigma$, at different values of the amplitude $I_{\text{max}}\in[0.8-1.5]$, in the absence of noise fluctuations. The other parameters are: $r=0.1$, $\gamma=0.25$, $\varepsilon =10$, $\omega=1$, and $m_z(t=0)=+1$.} 
\label{Figure03}
\end{figure}

For small enough $r$ values, the magnetization reversal effect is absent.
Interestingly, the pattern of dark fringes evidently changes for $I_{\text{max}}>1$, so that 
for $I_{\text{max}}=1.6$ the dark fringes merge together in large areas of $(\gamma,r)$ values in which magnetization reversal takes place.

With the aim of selecting a driving pulse suitable for a memory application, we observe that the magnetization reversal effect should be sufficiently robust against small changes of both the current pulse intensity and the duration of the pulse. In Fig.~\ref{Figure03} we show the stationary magnetization as a function of the pulse width $\sigma$, by changing the pulse amplitude $I_{\text{max}}$. We have chosen the parameters $r=0.1$ and $\gamma=0.25$. We observe that for a bias current below the critical value, the need for an accurate current-width regulation is significantly relaxed, since the magnetization reversal definitively occurs for any width above a specific value. Instead, for current amplitudes higher than the critical value, the stationary magnetization versus $\sigma$ is highly scattered between the two possible values, $m_z^{\text{st}}=\pm1$, so that even a slight change in the pulse width may lead to a non-switching situation. To understand this behavior, we observe that, for a bias current higher than $I_c$, a pulse sufficiently long can make the junction to switch to the resistive state, so that the Josephson phase rapidly evolves and a voltage drop across the device appears, since $V\propto\frac{\mathrm{d} \varphi}{\mathrm{d} t}$~\cite{Bar82}. In this case, the steady magnetization strongly depends on the dynamical state of the system when the current pulse is switched off. Moreover, the higher the bias current is, the faster the Josephson phase evolve and more pronounced will be the $\sigma$-dependence of $m_z^{\text{st}}$. In view of a memory application, a current pulse smaller than the critical value is therefore recommended, in order to make the magnetization reversal unrelated on the pulse width $\sigma$. For these reasons in the subsequent analysis we set $I_{\text{max}}=0.9$ and $\sigma=5$ and focus on the thermal effect on the phase and the magnetization dynamics.

\section{Effects of noise}
\label{noiseeffects}\vskip-0.2cm

In this section, we focus on the noisy dynamics of the junction, specifically on how it affects the magnetization reversal. The temperature can significantly influence the time evolution of the system, eventually inducing unwanted magnetization flip or preventing a stable magnetization reversal. Therefore, we consider stochastic thermal fluctuations in both the phase and the magnetic moment dynamics. We start discussing the effect of a thermal noise source only on the phase within the RSJ model. In the second part of this section we include also the thermal noise on the magnetization dynamics.

\subsection{Thermal current effects on the RSJ model}
\label{noiseRSJ}\vskip-0.2cm

\begin{figure}[t!!]
\centering
\includegraphics[width=\columnwidth]{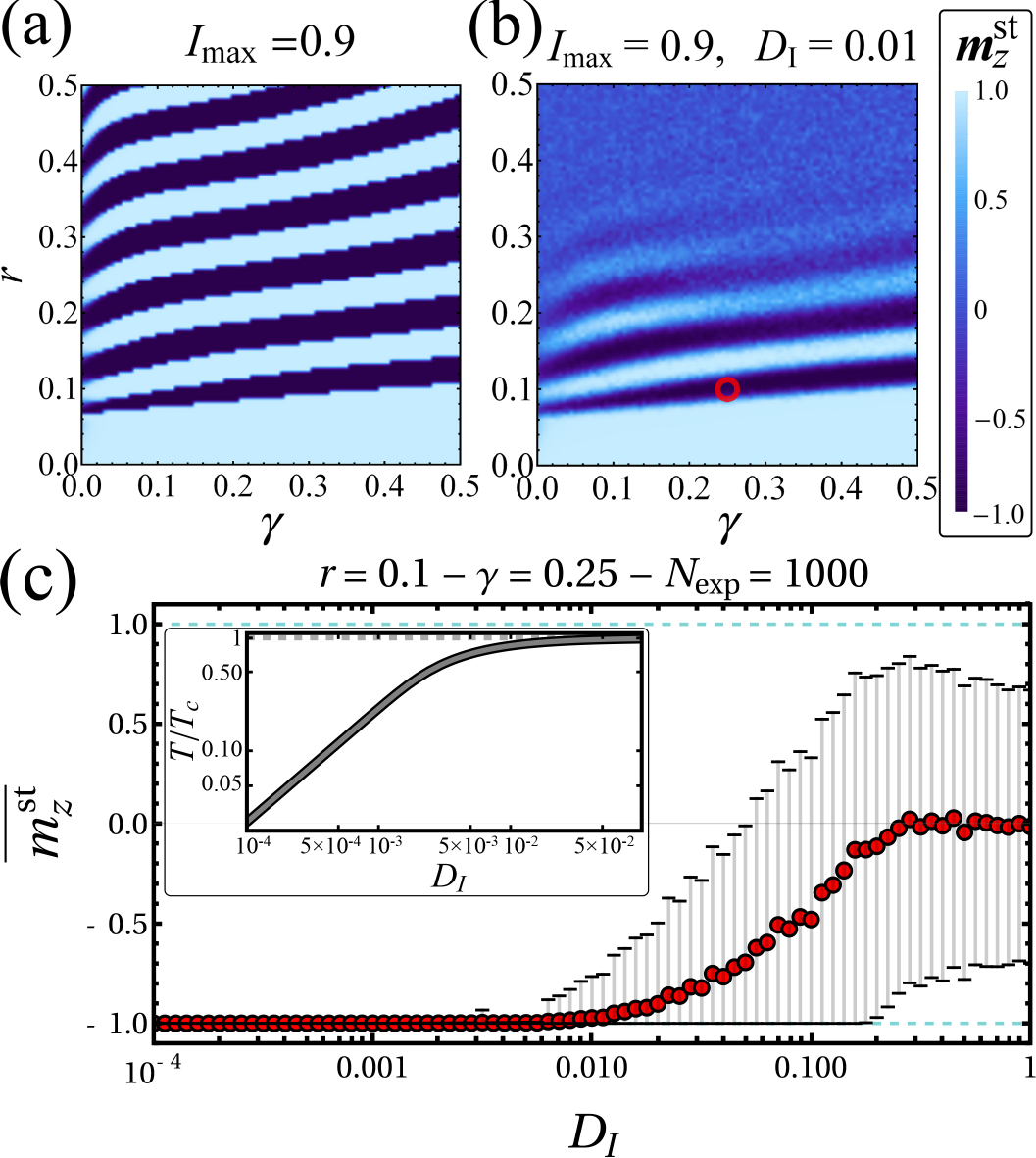}
\caption{(a) Stationary magnetization, $m_z^{\text{st}}$, as a function of $r$ and $\gamma$. (b) Average stationary magnetization, $\overline{m_z^{\text{st}}}$, as a function of $r$ and $\gamma$, at $\DI =0.01$, calculated by averaging over $N_{\text{exp}}=100$ independent numerical repetitions. (c) Average stationary magnetization, $\overline{m_z^{\text{st}}}$, as a function of the thermal current intensity, $\DI$, at $\gamma=0.25$ and $r=0.1$ [namely, the $(\gamma,r)$-values highlighted with a red circle in panel (b)], calculated by averaging over $N_{\text{exp}}=1000$ independent numerical repetitions. The inset shows the normalized temperatures corresponding to the noise intensities $\DI$. For all panels $I_{\text{max}}=0.9$ and $\sigma=5$, whereas the values of other parameters are the same used to obtain Fig.~\ref{Figure02}.} 
\label{Figure04}
\end{figure}

The phase dynamics can be directly perturbed by thermal fluctuations accounted by adding to the RSJ model, Eq.~\eqref{RSJ}, a Langevin Gaussianly distributed, delta correlated stochastic term, $I_{\text{th}}(t)$. This ``thermal current'' has the usual white-noise statistical properties that, in normalized units, can be expressed as~\cite{Bar82,Gua13,Val14}
\begin{eqnarray}
\left \langle I_{\text{th}}(t) \right \rangle&=&0\\
\left \langle I_{\text{th}}(t)I_{\text{th}}({t}') \right \rangle&=&2\DI \delta \left ( t-{t}' \right ).
\label{thermalcorrelatorI}
\end{eqnarray}
Here, we introduced the dimensionless amplitude of thermal current fluctuations defined as
\begin{equation}\label{thermalcurrentamplitude}
\DI =\frac{k_BT}{R}\frac{\omega_F}{I_c^2}=\frac{1}{\omega}\frac{k_BT}{ E_J}.
\end{equation}
For example, if $\omega=1$ we obtain $\DI \sim0.04\frac{T}{I_c}\frac{\mu\text{A}}{K}$, so that, for instance, a junction with $I_c=1\;\mu\text{A}$ at $T=250\;\text{mK}$ is affected by a thermal fluctuation of intensity $\DI \sim10^{-2}$.

By taking into account the noise contribution, Eq.~\eqref{RSJ} becomes
\begin{equation}\label{RSJnoise}
\frac{d\varphi}{dt}=\omega\left [ I_{bias}(t)-\sin\left ( \varphi-rm_y \right )+I_{\text{th}}(t)\right ]+ r\frac{d m_y}{dt}.
\end{equation}

In Fig.~\ref{Figure04} we compare the current-induced magnetization reversal obtained without and with accounting of the noise effects, see panel (a) and (b) respectively. We set the intensity of the current pulse $I_{\text{max}}=0.9$ and its width $\sigma=5$. 

In Fig.~\ref{Figure04}(a) we show the behavior of $m_z^{\text{st}}$ as a function of $r$ and $\gamma$ in the deterministic case, namely, in the absence of noise, $\DI=0$. Here, we observe a contour plot composed by many narrow dark fringes in which $m_z^{\text{st}}=-1$, see Fig.~\ref{Figure04}(a).

The situation drastically changes if we include the thermal noise. 
In this case we focus on the average stationary magnetization, $\overline{m_z^{\text{st}}}$, which is computed by averaging the stationary magnetization over $N_{\text{exp}}=100$ independent numerical runs. The behavior of $\overline{m_z^{\text{st}}}$ as a function of $r$ and $\gamma$ for $\DI=0.01$ is illustrated in Fig.~\ref{Figure04}(b). At small values of $r$ the magnetization reversal is still absent, whereas noise mostly affects the regions with large $r$ where the averaged value of the magnetization is mainly distributed around zero. 
Nevertheless, one can still identify dark regions in which magnetization switching takes place. With a red circle we highlight in Fig.~\ref{Figure04}(b) the region around the point $(\gamma,r)=(0.25,0.1)$ where the magnetization takes the largest negative average magnetization $\overline{m_z^{\text{st}}}\simeq-1$. In other words, the region with the most robust switching. 

By increasing the noise intensity the switching process is suppressed, as shown in Fig.~\ref{Figure04}(c), at the optimal values $r=0.1$ and $\gamma=0.25$ in Fig.~\ref{Figure04}(b). In Fig.~\ref{Figure04}(c) the value of $\overline{m_z^{\text{st}}}$ is the average over $N_{\text{exp}}=1000$ independent numerical repetitions. In the inset we show the normalized temperatures corresponding to the noise intensities $\DI$, calculated by assuming a junction with a temperature-dependent critical current and $I_c=100\;\mu\text{A}$ at low temperatures~\footnote{In the case of weak proximity effect and large exchange field in F, the critical current temperature-dependence is proportional to $I_c(T)\propto\Delta(T)\tanh \left [ \Delta(T)/(k_BT) \right ]$, where $\Delta (T)$ is the superconducting gap, see for example Eq.~(2.37) of Ref.~\cite{Ber05}. Thus, to find the temperature corresponding to a given noise intensity, we use this relation, with a zero-temperature value $I_c(0)=100\;\mu\text{A}$.}.
From this figure, one sees that $\overline{m_z^{\text{st}}}\simeq-1$ only for $\DI\lesssim0.01$, that is for $T\lesssim0.75T_c$. For higher noise intensities both the average magnetization and the error bar increase, approaching a zero magnetization average only for $\DI\gtrsim0.3$.

\begin{figure}[t!!]
\centering
\includegraphics[width=\columnwidth]{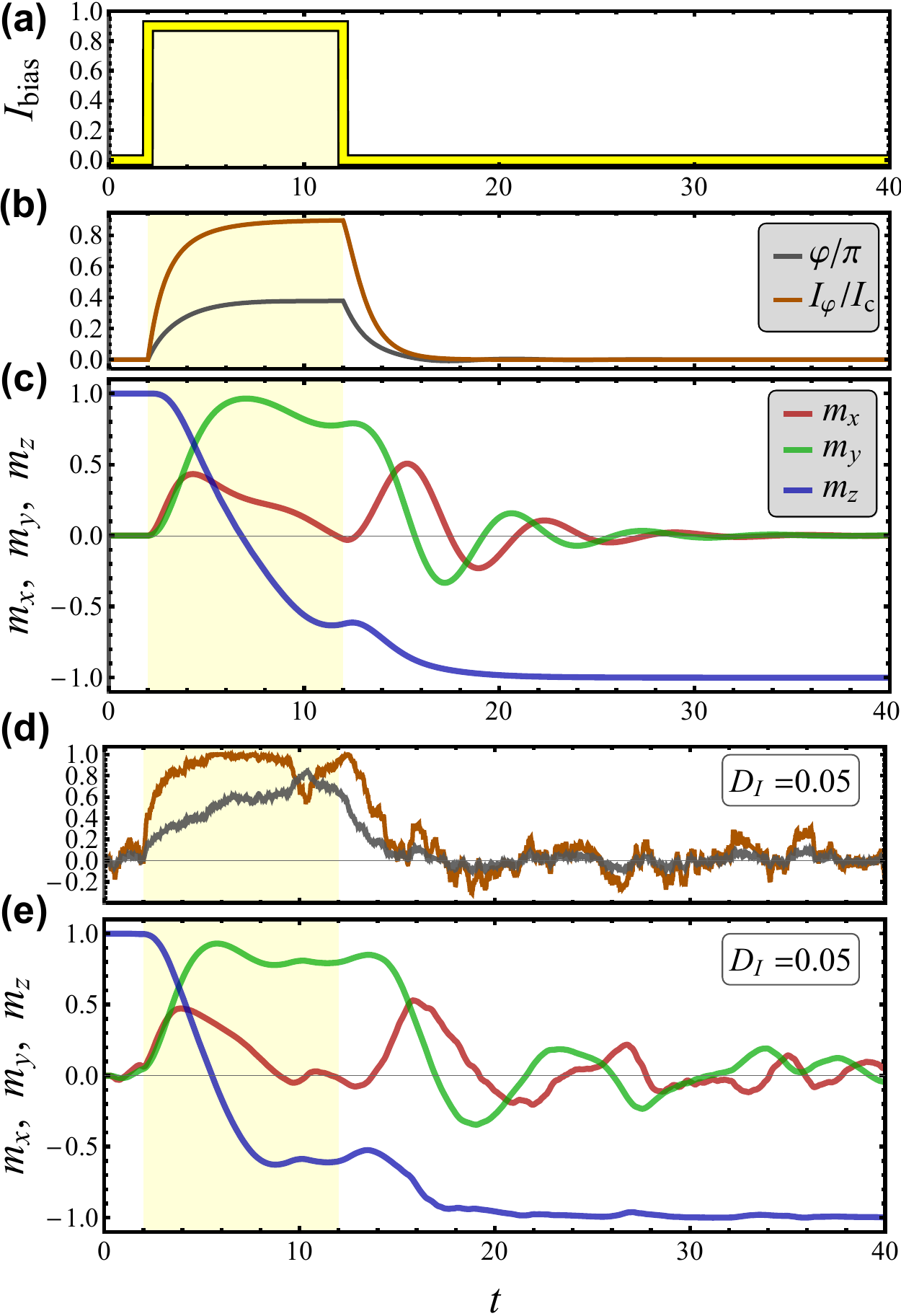}
\caption{Current pulse (a) and following time evolution of phase and Josephson current, see panel (b), and magnetization components, see panel (c), in the absence of noise and including a thermal current contribution with amplitude $\DI =0.05$, see panels (e) and (d). The values of other parameters are $I_{\text{max}}=0.9$, $\sigma=5$, $r=0.1$, and $\gamma=0.25$. The legends in panels (b) and (c) refer also to panels (d) and (e), respectively. } 
\label{Figure05}
\end{figure}
In Fig.~\ref{Figure05} we explore the time evolution of the different observables with and without thermal noise. 
Specifically, we show the response of the junction with $\gamma=0.25$ and $r=0.1$ to a current pulse with amplitude $I_{\text{max}}=0.9$ depicted in panel (a). In the absence of noise, $\DI=0$, we plot in Fig.~\ref{Figure05}(b) the time evolution of the phase and the supercurrent, and in Fig.~\ref{Figure05}(c) the different components of the magnetic moment.
During the current pulse, {\it i.e.}, the yellow shaded region, the phase first increases, and then it goes to zero when the pulse is turned off, see Fig.~\ref{Figure05}(b). To understand the phase behavior, we observe that, in the washboard-like picture~\cite{Bar82}, the tilting imposed by the bias current $I_{\text{max}}=0.9$ is not enough for allowing the ``particle'' to overcome the nearest potential barrier and switch the system to the finite voltage ``running'' state. Instead, the phase-particle remains confined within a potential minimum, so that when the current is turned off, the slope of the washboard potential goes again to zero and the phase restores its initial position, {\it i.e.}, $\varphi\to0$. 
 
We observe that the larger the bias current pulse is, the higher is the washboard potential slope, and therefore for $I_{\text{max}}>1$ the greater the speed of the phase particle, so that it can take a longer time to restore the initial position after the current pulse is switched off. Moreover, a large bias current pulse may also longer switching times. Hence, a current $I_{\text{max}}<1$ is, in general, more advantageous for a memory application.

In Fig.~\ref{Figure05}(c) we show how all components of the magnetization are induced by the current pulse. Whereas $m_x$ and $m_y$ are generated during the current pulse, and they undergo a damped oscillations around zero when the current is switched off, the $z$-component, after a transient regime, flips definitively to the value $m_z=-1$. From this figure we can also estimate the switching time $t_{\text{SW}}\simeq10$, as the time $m_z$ roughly takes to approach the value $-1$ after switching off the current. 

The scenario described so far essentially persists also in the stochastic case, as shown in Figs.~\ref{Figure05}(d-e) for $\DI=0.05$. Therefore, at the temperature that we are considering, the overall behavior is still quite similar to the one obtained in the absence of noise. 
In fact, the $z$-component of the magnetization flips again to the value $-1$, while the $x$ and $y$ components tend to oscillate around zero, without, however, vanishing definitively.

\begin{figure}[t!!]
\centering
\includegraphics[width=\columnwidth]{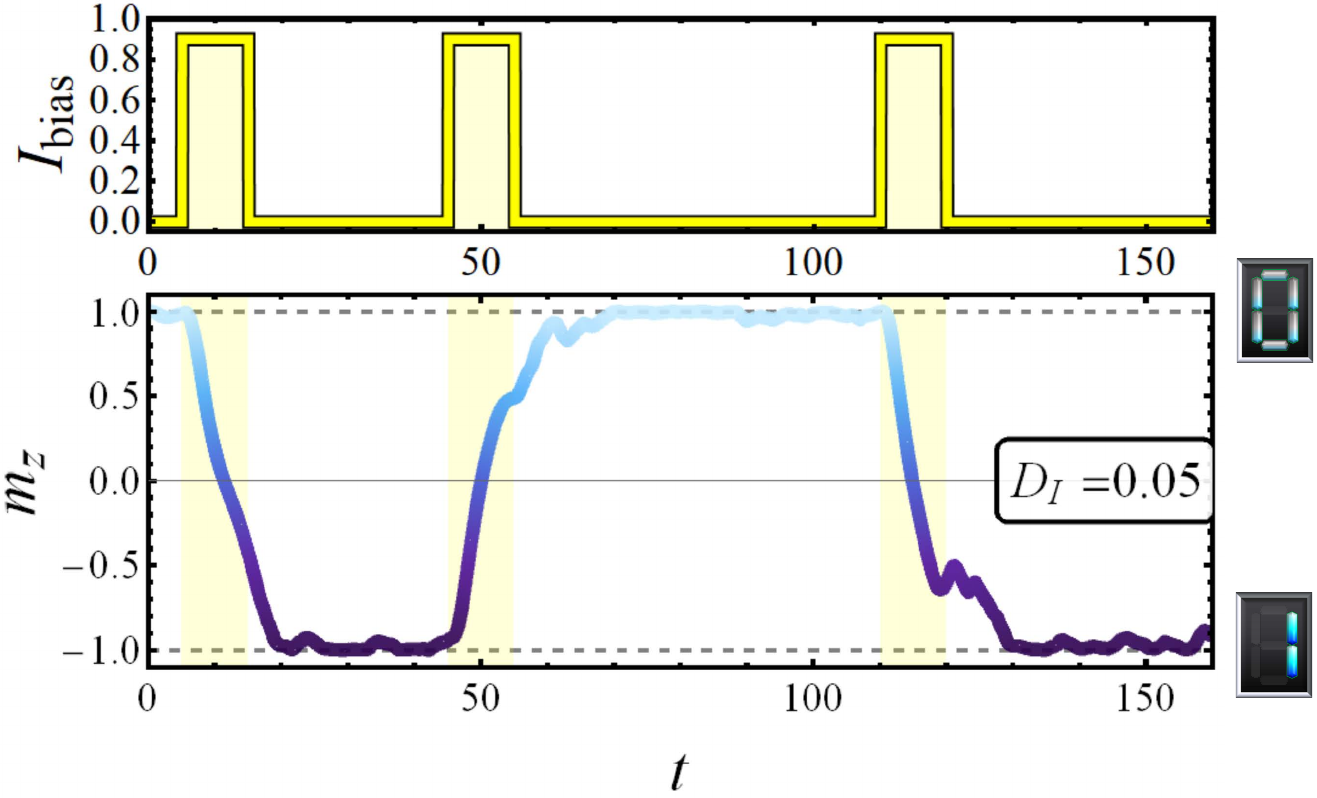}
\caption{Time evolution of the magnetization $m_z$, in response to a sequence of three current pulses shown in the top panel, in the presence of a thermal current noise with amplitude $\DI =0.05$. The values of the other parameters are: $I_{\text{max}}=0.9$, $\sigma=5$, $r=0.1$, and $\gamma=0.25$.} 
\label{Figure06}
\end{figure}

The magnetization switch can be achieved in a short time scale, by passing through the junction a sequence of current pulses, as it is shown in Fig.~\ref{Figure06}. In the bottom panel we show the time evolution of the magnetization $m_z$, when the junction is excited by the three subsequent current pulses presented in the top panel, in the presence of a thermal current noise with amplitude $\DI =0.05$. In response to each current pulse, $m_z$ follows first a transient regime, and then, as the current is switched off, it approaches the steady value with an opposite sign.

\subsection{Effect of thermal noise on the magnetization dynamics }
\label{noiseLLG}\vskip-0.2cm

Thermal noise also affects directly the magnetization dynamics~\cite{Gar98,Sun06,Ber07,Cof12,Nis15,Lel17} via 
a stochastic field $H_{\text{th}}$, a sort of ``thermal field'', which is added 
to the effective magnetic field term in Eq.~\eqref{LLG}, as done in Ref.~\cite{Bro63}.
Inclusion of the thermal noise in Eq.~\eqref{LLG} leads to~\cite{Rom14}
\begin{equation}\label{LLGnoise}
\frac{d\textbf{M}}{d\tau}=-g_r\textbf{M}\times(\textbf{H}_{\text{eff}}+\textbf{H}_{\text{th}})+\frac{\gamma}{M}\left ( \textbf{M}\times\frac{d\textbf{M}}{d\tau} \right ).
\end{equation}
This stochastic differential equation has to be solved numerically by a stochastic integration prescription by keeping the modulus of the magnetic moment constant during the time evolution (see Ref.~\cite{Rom14} and references therein). For this purpose it is again convenient to write the equations in spherical coordinates, see Eq.~\eqref{Msphericalcoord}, so that the stochastic LLG equation reads ~\cite{Rom14,Aro14}:
\begin{eqnarray}\label{LLGsphericalcoord_Noise_a}
&&\frac{\mathrm{d} \theta}{\mathrm{d}t}=\frac{1}{1+\gamma^2}\Big [\widetilde{H}_{\text{eff},\phi}+\widetilde{H}_{\text{th},\phi} +\gamma\left ( \widetilde{H}_{\text{eff},\theta}+\widetilde{H}_{\text{th},\theta}\right ) \Big ]\\\label{LLGsphericalcoord_Noise_b}
&&\sin \theta \frac{\mathrm{d} \phi}{\mathrm{d} t}\!=\!\frac{1}{1+\gamma^2}\Big [\gamma\left ( \widetilde{H}_{\text{eff},\phi}\!+\! \widetilde{H}_{\text{th},\phi}\right )\!-\! \widetilde{H}_{\text{eff},\theta}\!-\! \widetilde{H}_{\text{th},\theta} \Big ], 
\end{eqnarray}
where
\begin{eqnarray}\label{AngleFieldsThermal}
&&\widetilde{H}_{\text{th},\theta}\!=\!\widetilde{H}_{\text{th},x}\cos\theta\cos\phi\!+\!\widetilde{H}_{\text{th},y}\cos\theta\sin\phi\!-\!\widetilde{H}_{\text{th},z}\sin\theta\;\;\;\;\;\;\\
&&\widetilde{H}_{\text{th},\phi}\!=\!-\!\widetilde{H}_{\text{th},x}\sin\phi\!+\!\widetilde{H}_{\text{th},y}\cos\phi.
\end{eqnarray}
\begin{figure}[t!!]
\centering
\includegraphics[width=\columnwidth]{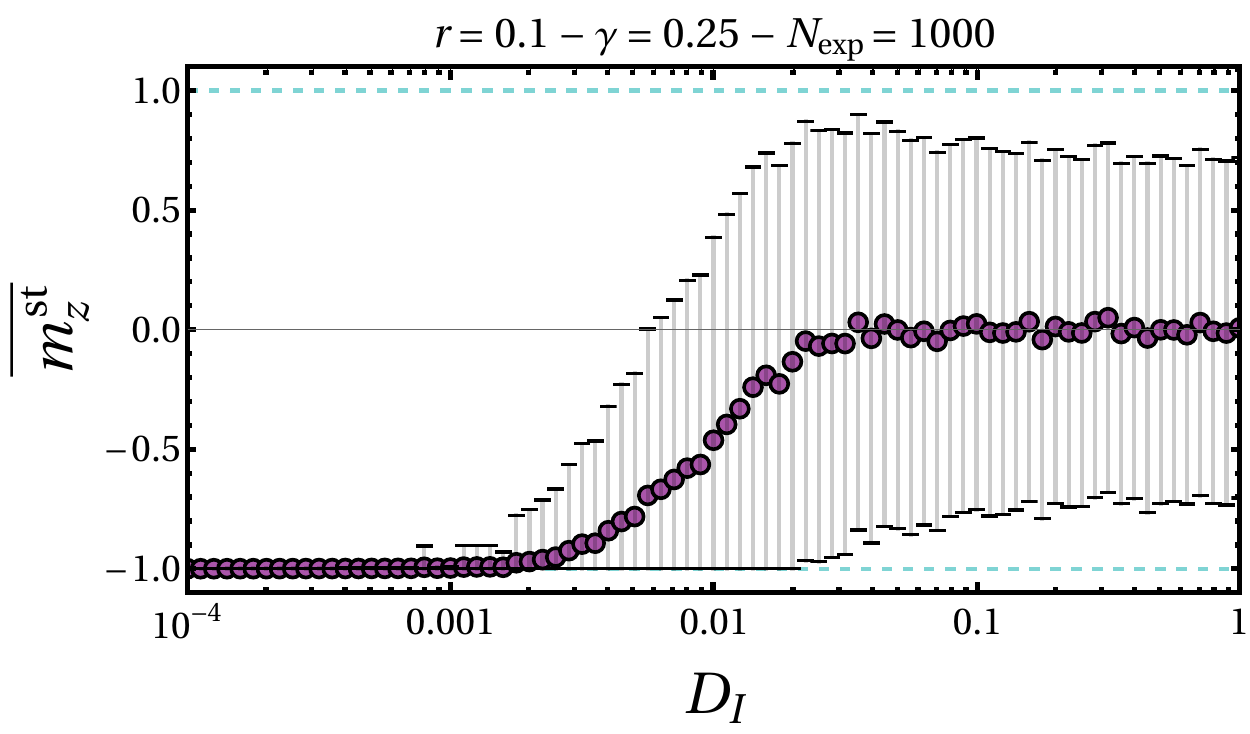}	
\caption{Average stationary magnetization, $\overline{m_z^{\text{st}}}$, as a function of the noise intensity, $\DI $, calculated by taking into account both the thermal current and the thermal field noise contribution, and by averaging over $N_{\text{exp}}=1000$ independent numerical repetitions. The values of other parameters are $I_{\text{max}}=0.9$, $\sigma=5$, $r=0.1$, and $\gamma=0.25$.} 
\label{Figure07}
\end{figure}
The normalized field, $\widetilde{\textbf{H}}_{\text{\text{th}}}=(M/\mathcal{K})\textbf{H}_{\text{th}}$ is assumed to be a Gaussianly distributed random field with the following statistical features
\begin{eqnarray}
\left \langle \widetilde{H}_{\text{th},i}(t) \right \rangle&=&0\\
\left \langle \widetilde{H}_{\text{th},i}(t)\widetilde{H}_{\text{th},i}({t}') \right \rangle&=&2\Dh \delta \left ( t-{t}' \right ),
\label{thermalcorrelatorH}
\end{eqnarray}
where $i=x,y,z$ and
\begin{equation}\label{thermalfieldamplitude}
\Dh =\left(\frac{\gamma }{M}\frac{k_BT}{|g_r|\Omega}\right)\left(\frac{M}{\mathcal{K}}\right)^{\!2}\!\!\!\omega_F=\gamma\frac{k_BT}{\mathcal{K}\Omega}
\end{equation}
is the dimensionless amplitude of thermal field fluctuations. In all previous equations the time is still normalized to the inverse of $\omega_F$. 

Interestingly, by recalling the definition of the parameter $\varepsilon =E_J/(\mathcal{K}\Omega)$, from Eqs.~\eqref{thermalcurrentamplitude} and~\eqref{thermalfieldamplitude} we can easily obtain the following relation between the normalized thermal noise intensities
\begin{equation}\label{noiserelation}
\Dh =(\gamma\, \varepsilon \omega)\DI .
\end{equation}
Thus, by changing the magnetization energy, the Gilbert damping parameter, or the magnetic resonance frequency we can effectively modify the relative strength of the two noise mechanisms. This means that one could optimize the system parameters in such a way to make, for instance, the impact of the thermal field negligible with respect to the thermal current. This allows us to study the effects produced by these noise sources independently. In the following, even if we explicitly write only the value of $\DI$, we are taking into account both thermal current and thermal field independent noise sources, which amplitudes are related by Eq.~\eqref{noiserelation}.

The overall effect of both the thermal current and field is presented in Fig.~\ref{Figure07}, where we show the behavior of $\overline{m_z^{\text{st}}}$, calculated by averaging over $N_{\text{exp}}=1000$ independent numerical runs, at different values of the noise intensity $\DI$, and by setting $I_{\text{max}}=0.9$, $\sigma=5$, $\gamma=0.25$, and $r=0.1$. We observe that the average magnetization remains close to the value $\overline{m_z^{\text{st}}}\simeq-1$ only for $\DI\lesssim0.003$, that is for $T\lesssim0.58T_c$, see the inset of Fig.~\ref{Figure04}(c). For larger values of $\DI$, $\overline{m_z^{\text{st}}}$ approaches zero and hence the magnetization reversal probability is reduced, Fig.~\ref{Figure04}(c). 
In view of the memory application, one should, in principle, carefully choose the F layer and its characteristics (such as its volume or the Gilbert damping parameter) in order to make the thermal field effect as small as possible. The aim is to reduce the thermal field intensity in order to increase the working temperature suitable for a memory application, {\it e.g.}, through a lower Gilbert damping or a larger F volume, according to Eq.~\eqref{noiserelation}. 

\begin{figure}[t!!]
\centering
\includegraphics[width=\columnwidth]{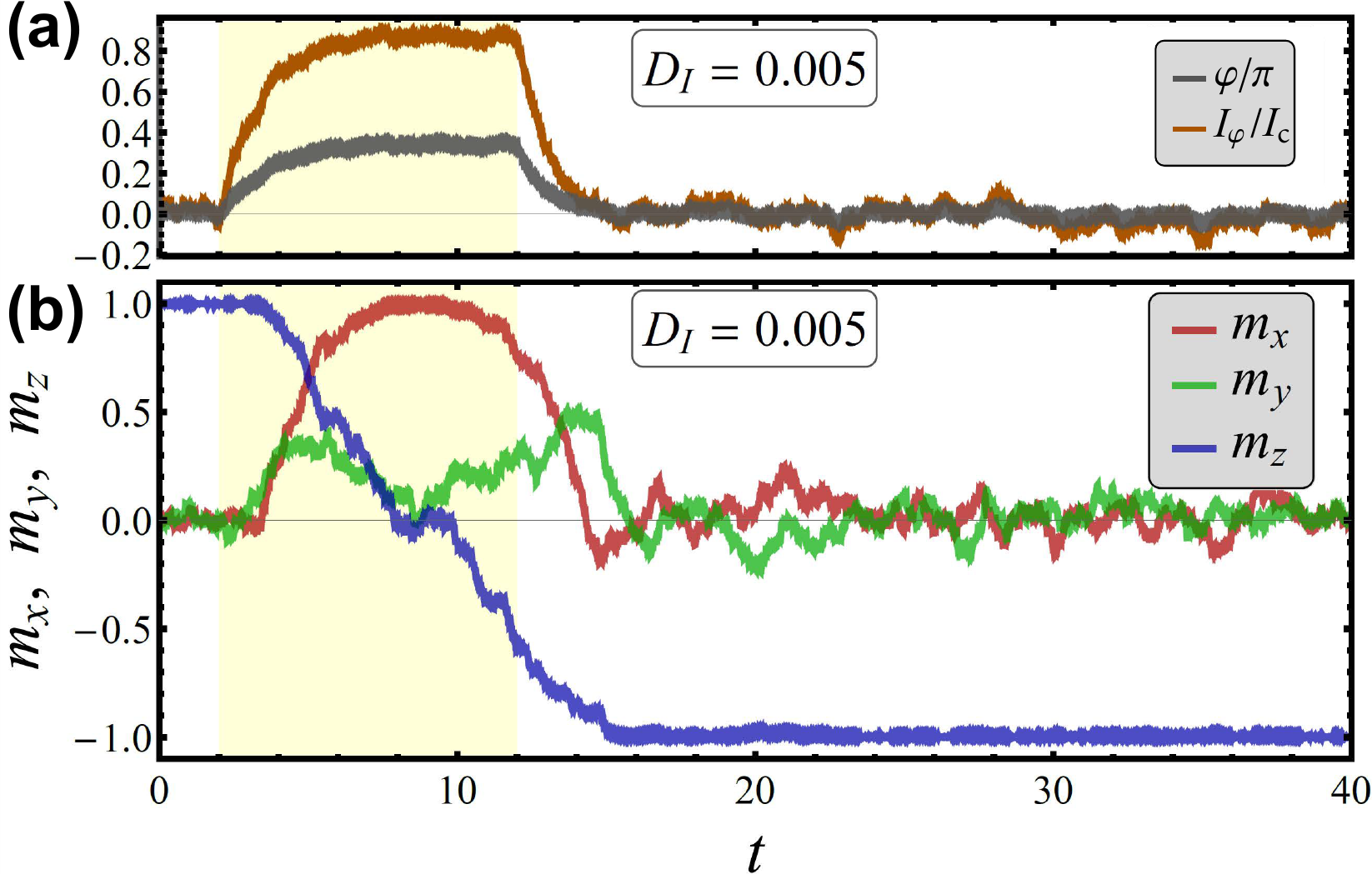}	
\caption{Time evolution of phase and Josephson current (a) and magnetization components (b) as the system is excited by the current pulse in Fig.~\ref{Figure05}. Here we are taking into account both a thermal current and a thermal field contribution, with noise intensity $\DI =0.005$. The values of other parameters are $I_{\text{max}}=0.9$, $\sigma=5$, $r=0.1$, and $\gamma=0.25$.} 
\label{Figure08}
\end{figure}

The time evolution of $\varphi$, $I_{\varphi}$, and $m_i$ (with $i=x,y,z$), as the junction dynamics if affected by both a thermal current and a thermal field, for $r=0.1$, $\gamma=0.25$, and $\DI =0.005$, is shown in Fig.~\ref{Figure08}. Here, we consider again the system excited by a current pulse with intensity $I_{\text{max}}=0.9$, as that one shown in Fig.~\ref{Figure05}(a). We observe that all noisy curves still resemble in shape the deterministic evolution presented in Figs.~\ref{Figure05}(b)-(c). The value $t_{\text{SW}}\simeq10$ is a quite good estimation for the switching time of the device also in this noisy case.

\section{Rashba-Dresselhaus SOC}
\label{GenericSOC}\vskip-0.2cm

 In all previous analysis it was assumed a pure Rashba SOC. However, the theory of $\varphi_0$-junctions can be generalized for any linear-in-momentum SOC~\cite{Ber15,Kon15}, by using the SU(2)-covariant formulation~\cite{Ber15}, where the SOC is described in terms of a SU(2) vector potential $\mathcal{A}$. 
 For a 2D SOC with both Rashba and Dresselhaus contributions one obtains
 $\mathcal{A}_x=-\alpha\sigma^y+\beta\sigma^x$ and $\mathcal{A}_y=\alpha\sigma^x-\beta\sigma^y$~ (here, $\alpha$ and $\beta$ are the Rashba and Dresselhaus coefficients and $\sigma^x$ and $\sigma^y$ are the first two Pauli matrices).
 
The appearance of the anomalous phase is related to the existence of a finite Liftshitz invariant term in the free energy~\cite{kaur2005helical,samokhin2004magnetic,mineev2011magnetoelectric,Buz08} which is proportional to $T_i\partial_i\varphi$, where $T_i$ is the $i$-th component of a polar vector which is odd under time reversal, $\partial_i$ is the $i$-the derivative of the superconducting phase, and the sum over repeated indices is implied here and below. For the particular junction geometry sketched in Fig.~\ref{Figure01} the supercurrent, and hence the phase gradient, is finite in $x$-direction. Thus, according to Eq.(5.17) of Ref.~\cite{Kon15}, the anomalous phase can be written in the following compact form~\cite{Ber15,Kon15}:
 \begin{equation}
\label{phi0genericSOC}
\varphi_0=\rb (\wb m_x+m_y).
\end{equation}
Here, we defined the SOC coefficients ratio, $\wb=\beta/\alpha$, and the parameter $\rb =r(1-\wb^2)$, with $r$ depending this time on both $\alpha$ and $\beta$. 
In the absence of the Dresselhaus SOC, that is when $\wb=0$ and $\rb\to r$, we recover Eq.~\eqref{phi0}. 
If both contributions are similar in magnitude, {\it i.e.}, when $\wb\to1$, since $\rb \to0$ the phase shift vanishes, {\it i.e.}, $\varphi_0\to0$. This is a very interesting situation that we explore in this section. In fact, whereas the Dresselhaus 
contribution is due to the breaking of crystal inversion symmetries, the Rashba SOC stems from structural broken symmetry and therefore can be controlled by a gate voltage~\cite{Duc09,Det17}. In other words, a voltage gate can control the ratio $\wb$ between Dresselhaus and Rashba coefficients, and hence the phase shift and the supercurrent flow, according to Eq.~\eqref{phi0genericSOC}. Specifically, by tuning $\alpha$ such that $\wb\simeq1$ one can fully decouple the phase and magnetic moment dynamics. Such a process can be eventually used to protect the memory state in one of the storage elements of a distributed architecture.

\begin{figure}[t!!]
\centering
\includegraphics[width=\columnwidth]{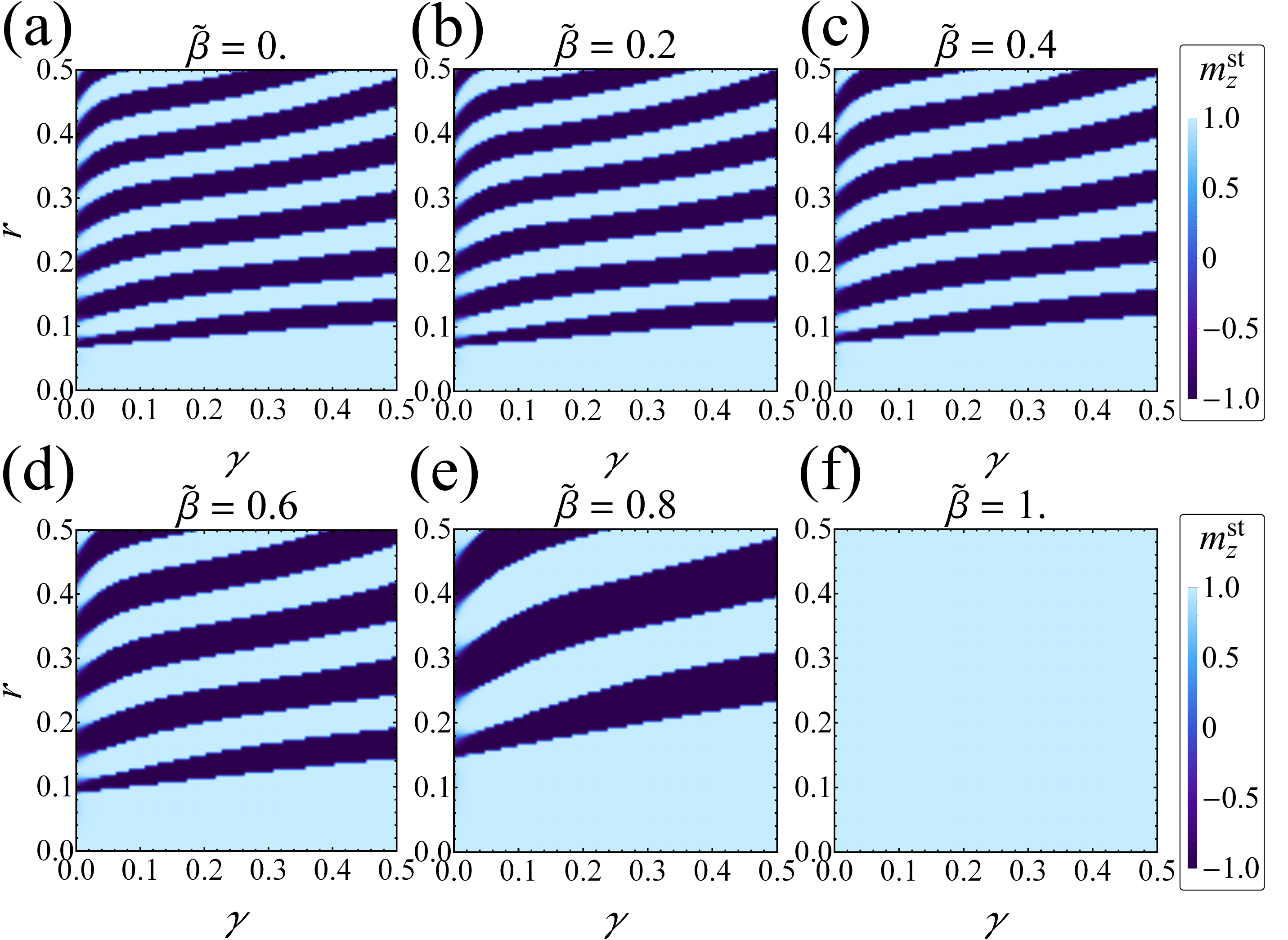}
\caption{Stationary magnetization, $m_z^{\text{st}}$, as a function of $r$ and $\gamma$, at different values of the relative Dresselhaus coefficient $\wb=\beta/\alpha$, in the absence of noise fluctuations, by imposing $I_{\text{max}}=0.9$ and $\sigma=5$.} 
\label{Figure09}
\end{figure}

We provide next a quantitative analysis of this situation, so that by taking into account the generic $\varphi_0$, Eq.~\eqref{phi0genericSOC}, into the expression for the effective field, Eq.~\eqref{effectivefield} becomes
%
%
%
\begin{equation}\label{effectivefielddress}
\textbf{H}_{\text{eff}}=\frac{\mathcal{K}}{M}\left \{\varepsilon\rb \sin\left[\varphi-\rb \left(\wb m_x+m_y\right)\right]\left(\wb\hat{x}+\hat{y}\right)+m_z\hat{z}  \right\}\!. 
\end{equation}
The $\theta$ and $\phi$ components of the normalized effective field to be included in LLG Eqs.~\eqref{LLGsphericalcoord_a}-\eqref{LLGsphericalcoord_b}, read 
%
%
\begin{eqnarray}\label{AngleFieldsEffective}\nonumber
\widetilde{H}_{\text{eff},\theta}=&&\varepsilon \rb\cos\theta\sin\left[\varphi-\rb \left(\wb m_x+m_y\right)\right]\left(\wb\cos\phi+\sin\phi\right)\\
&&-m_z\sin\theta\\
\widetilde{H}_{\text{eff},\phi}=&&\varepsilon \rb\sin\left[\varphi-\rb \left(\wb m_x+m_y\right)\right]\left(\cos\phi-\wb\sin\phi\right) ,
\end{eqnarray}
whereas the RSJ equation becomes
\begin{eqnarray}\label{RSJ_genericSOC}
\frac{d\varphi}{dt}=&&\omega\left\{I_{bias}(t)-\sin\left[\varphi-\rb \left(\wb m_x+m_y\right)\right]\right\} \\\nonumber
&&+ \rb \left(\wb\frac{d m_x}{dt}+\frac{d m_y}{dt}\right).
\end{eqnarray}
The behavior of the stationary magnetization as a function of $r$ and $\gamma$, at different values of $\wb\in[0-1]$ is shown in Fig.~\ref{Figure09}. The current pulse intensity and width are chosen equal to $I_{\text{max}}=0.9$ and $\sigma=5$, respectively. As expected from the discussion above, the region
where no magnetization switching occurs, bright color in Fig.~\ref{Figure09}, increases by increasing $\wb$ towards $1$. 
For $\wb=1$, the $\varphi_0$ behavior is fully suppressed, {\it cf. }Eq.~\eqref{phi0genericSOC}, and hence no magnetization switching takes place, despite the current pulse flowing through the junction. Interestingly, we note that for intermediate values of $\wb$, the area of the switching fringes, {\it i.e.}, where $m_z^{\text{st}}=-1$, increases considerably.
 
In summary of this section, by tuning $\alpha=\beta$ one can decouple the magnetic behavior from the Josephson dynamics and freeze the memory state
in order to protect it from external current pulses and other perturbations. The major challenge in this regard is to find materials with a sizable 
magnetic moment and tunable by means of a gate voltage.

\section{The memory read-out}
\label{Readout}\vskip-0.2cm

As discussed in previous section, the writing operation of the proposed memory element can be performed by exciting the junction with controlled current pulses. We could envisage an array of $\varphi_0$-junction-based memory elements, each one eventually having its own current line so that it can be written by sending individual current inputs. Alternatively, exploiting the tuning of the SOC discussed in previous section, one could control locally, via individual gates at each junction, several memory elements connected in series to a common current line. In this way one could selectively write via a common current pulse only a specific set of memory units.

The read-out of the memory state can be non-destructively performed by direct measurement of the magnetization state through a dc-SQUID inductively coupled to the $\varphi_0$-junction. A SQUID is essentially a magnetic flux detector~\cite{Cla04}, which can be employed to measure with a very high sensitivity any physical quantity that can be converted in a magnetic flux~\cite{Gra16}.

\begin{figure}[t!!]
\centering
\includegraphics[width=\columnwidth]{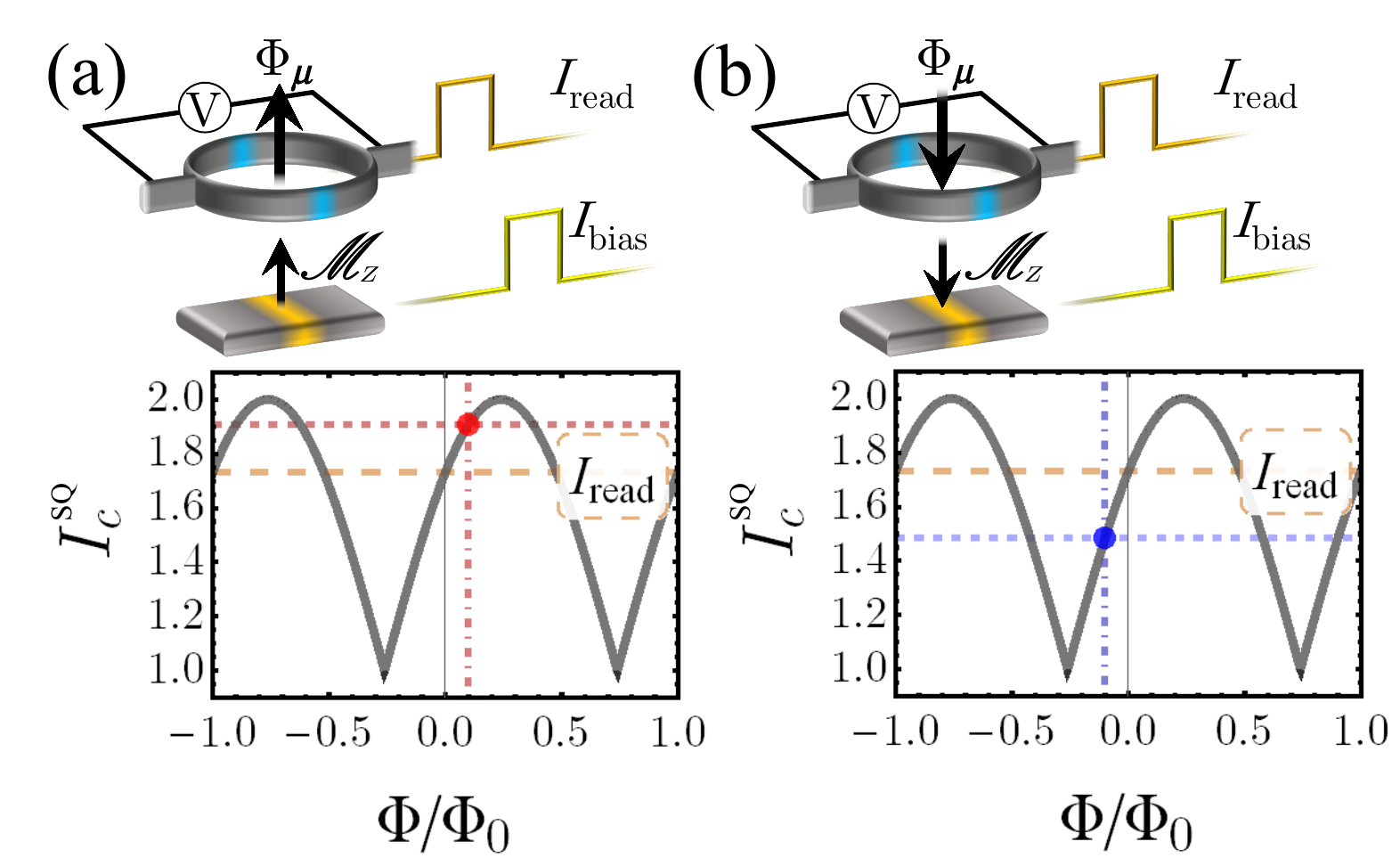}
\caption{SQUID-based memory readout and cartoon showing the critical current interference pattern of the SQUID, in the cases of both positive and negative orientation along the $z$-axis of the magnetic moment, see panel (a) and (b), respectively.} 
\label{Figure10}
\end{figure}

We suggest a SQUID sensor along the lines of the readout scheme implemented in Ref.~\cite{Day18} for a $\pi$-junction memory. Our scheme is based on an asymmetric inductive dc-SQUID, which consists of a superconducting ring with a non-negligible total inductance, $L$, with two Josephson junctions with different critical currents, {\it i.e.}, $I_{c,1}\neq I_{c,2}$ (here, we are neglecting for simplicity any asymmetry in the ring inductance). With such asymmetric SQUID, one can avoid the use of an additional magnetic flux to adjust the working point of the device in a high sensitivity position of the $I^{\textsc{SQ}}_c-\Phi$ characteristics, where $I^{\textsc{SQ}}_c$ is the SQUID critical current and $\Phi$ is the magnetic flux threading the loop. In fact, such asymmetric dc-SQUID shows non-negligible screening and asymmetry parameters, that is $\beta_L=\frac{2\pi }{\Phi_0}\frac{L}{2}(I_{c,1}+I_{c,2})\neq0$ and $\alpha_I=\frac{I_{c,1}-I_{c,2}}{I_{c,1}+I_{c,2}}\neq0$, respectively. In this case, the $I^{\textsc{SQ}}_c$ maximum is not centered in $\Phi=0$, but shifted from zero by $\Delta \Phi$, see Fig.~\ref{Figure10}, where~\cite{Cla04} $2\pi\Delta \Phi=\Phi_0\beta_L\alpha_I$. 
Accordingly, our readout SQUID demonstrates a high sensitivity point of the $I^{\textsc{SQ}}_c-\Phi$ characteristics also in $\Phi=0$, that is in the absence of an external magnetic flux. 

 We assume that the unbiased critical current, $I^0_c=I^{\textsc{SQ}}_c(\Phi=0)$, lies in the positive branch of the critical current diffraction pattern of the SQUID, that is $\left . \frac{\mathrm{d} I^{\textsc{SQ}}_c}{\mathrm{d} \Phi} \right |_{\Phi=0}>0$, just like in the case shown in Fig.~\ref{Figure10}. Thus, the magnetic moment $m_z=+1$ generates a positive magnetic flux $\Phi=+\Phi_{\mu}$ through the loop, and gives a critical current higher than the unbiased value, {\it i.e.}, $I_c^+=I^{\textsc{SQ}}_c(+\Phi_{\mu})>I^0_c$, see the red dashed line in Fig.~\ref{Figure10}(a). Conversely, if $m_z=-1$ the magnetic flux through the loop is negative, {\it i.e.}, $\Phi=-\Phi_{\mu}$, and the critical current is lower than the unbiased value, $I_c^-=I^{\textsc{SQ}}_c(-\Phi_{\mu})<I^0_c$, see the blue dashed line in Fig.~\ref{Figure10}(b).

The SQUID readout loop is then sensed by passing a bit-read current with intensity $I_{read}=I^0_c$ through the device, see the orange dashed line in Fig.~\ref{Figure10}, that is a current which lays in between $I_c^-$ and $I_c^+$. In this way, if the magnetic moment points in the negative $z$-direction, which encodes the ``1'' logic state of the memory, the bit-read current makes the SQUID to switch to the voltage state, since $I_{read}>I_c^-$. Consequently, a voltage drop appears across the readout SQUID in response to the bit-read current. For the opposite magnetic moment orientation, which encodes the ``0'' logic state, the SQUID critical current is larger than the bit-read current, that is $I_{read}<I_c^+$, so that the SQUID remains in the superconducting, zero-voltage state.

To provide an estimate of the dimensions of a SQUID loop suitable to detect the magnetic moment reversal we consider the simple case of a circular readout SQUID with radius $a$ and a magnetic moment oriented along the $z$-axis and placed on the symmetry axis of the SQUID at a height ${z}'$ above the $x-y$ plane. In this case, the total magnetic flux through the loop can be estimated as~\cite{Til09,Gra16} $\Phi_{\mu}=\frac{\mu_0M_z}{2}\;a^2\big/\left (a^2+{z}'^2 \right )^{3/2}$. This flux is maximum at ${z}'=0$ where it reads $\Phi_{\mu}=\frac{\mu_0M_z}{d}$, with $d=2a$ being the loop diameter. If we assume a saturation magnetization~\cite{Rus04} $M=8\times10^5\;\text{A}/\text{m}$ and a volume $\Omega=(10\times100\times100)\;\text{nm}^3=10^{-22}\;\text{m}^3$, we get a total magnetic moment $\mathcal{M}_z=8\times10^{-17}\;\text{A}\,\text{m}^2$ (that is $\mathcal{M}_z=8.6\times10^6\;\mu_\text{B}$, with $\mu_\text{B}$ being the Bohr magneton), so that the total magnetic flux reads $\Phi_{\mu}=32\pi10^{-24}d^{-1}\;\text{Wb}\;\text{m}$. 
Thus, for a SQUID with diameter $\simeq0.5\;\mu\text{m}$ we obtain a measurable flux $\Phi_{\mu}\simeq\Phi_0/10$, (see vertical dot-dashed lines in Fig.~\ref{Figure10}).

\section{Conclusions}
\label{Conclusions}\vskip-0.2cm
In conclusion, we discuss a non-volatile superconducting memory based on a bistable magnetic behavior of a current-biased $\varphi_0$-junction, that is a superconductor-ferromagnet-superconductor Josephson junction with a Rashba-like spin-orbit coupling (SOC). The memory state is encoded in the magnetization direction of the ferromagnetic layer, which can be switched via controlled current pulses.
Following Ref.~\onlinecite{Rab19}, the numerical approach that we used to describe the phase dynamics of a current-driven ferromagnetic Josephson junction is amended to include the time derivative of the anomalous phase.
We explore the robustness of the current-induced magnetization reversal against thermal fluctuations, in order to find the optimal working temperature at which the magnetization switching induced by a current pulse is stable. We also suggest a way of decoupling the Josephson phase and the magnetization dynamics, by tuning the Rashba SOC strength via a gate voltage.

Finally, we discuss a suitable non-destructive readout scheme based on a dc-SQUID inductively coupled to the $\varphi_0$-junction. The suggested readout scheme is exclusively controlled by current pulses, and no additional magnetic flux is needed.

\begin{acknowledgments}
This work was supported by the Horizon Research and Innovation Program under Grant Agreement No. 800923 (SUPERTED) 
and the Spanish Ministerio de Economía, Industria y Competitividad (MINEICO) under Project No. FIS2017-82804-P.
\end{acknowledgments}


%

\end{document}